\documentclass[aps,prb,twocolumn,superscriptaddress,showpacs]{revtex4}
\usepackage{graphicx, SIunits}
\usepackage[dvips]{color}

\begin{document}

\newcommand{\la}{\langle}
\newcommand{\ra}{\rangle}

\title{Acoustic Faraday effect in Tb$_3$Ga$_5$O$_{12}$}

\author{A.~Sytcheva}
\affiliation{Hochfeld-Magnetlabor Dresden (HLD), Forschungszentrum Dresden-Rossendorf, 01314 Dresden, Germany}

\author{U.~L\"ow}
\affiliation{
Technische Universit\"at Dortmund, 44221 Dortmund, Germany} 

\author{S.~Yasin}
\affiliation{Hochfeld-Magnetlabor Dresden (HLD), Forschungszentrum Dresden-Rossendorf, 01314 Dresden, Germany}

\author{J.~Wosnitza}
\affiliation{Hochfeld-Magnetlabor Dresden (HLD), Forschungszentrum Dresden-Rossendorf, 01314 Dresden, Germany}

\author{S.~Zherlitsyn}
\affiliation{Hochfeld-Magnetlabor Dresden (HLD), Forschungszentrum Dresden-Rossendorf, 01314 Dresden, Germany}

\author{P. Thalmeier}
\affiliation{
Max-Planck-Institut f\"ur Chemische Physik fester Stoffe, 01187 Dresden, Germany}

\author{T.~Goto}
\affiliation{
Graduate School of Science and Technology, Niigata University, 950-2181 Niigata, Japan}

\author{ P.~Wyder}
\affiliation{Laboratoire National des Champs Magn\'etiques Intenses, 38042 Grenoble Cedex 09, France}

\author{B.~L\"uthi}
\affiliation{Physikalisches Institut, Universit\"at Frankfurt, 60054 Frankfurt, Germany}


\begin{abstract}
The transverse acoustic wave propagating along the [100] axis of the cubic Tb$_3$Ga$_5$O$_{12}$ (acoustic $c_{44}$ mode) is doubly degenerate. A magnetic field applied in the direction of propagation lifts this degeneracy and leads to the rotation of the polarization vector - the magneto-acoustic Faraday rotation.  Here, we report on the observation and analysis of the magneto-acoustic Faraday-effect in Tb$_3$Ga$_5$O$_{12}$ in static and pulsed magnetic fields.  
We present also a theoretical model based on magnetoelastic coupling of 4$f$ electrons to both, acoustic and optical phonons and an effective coupling between them. This model explains the observed linear frequency dependence of the Faraday rotation angle.

\end{abstract}



\maketitle

\section{Introduction}

TGG  is a dielectric material with a cubic garnet structure. The orthorhombic local symmetry for Tb$^{3+}$ ($J=6$) leads to pronounced crystal electric field (CEF) effects and a Curie-type magnetic susceptibility. At $T_{{\rm N}} = \unit{0.35}{\kelvin}$ an antiferromagnetic transition appears. \cite{afmorder, afmorder1, afmorder2} 

Remarkable symmetry experiment, the so-called phonon Hall effect, was recently performed and analysed in Tb$_3$Ga$_5$O$_{12}$ (TGG) \cite{hall1, hall2, hall3, hall4}: in analogy to the classical Hall effect in conducting materials, the appearance of a thermal gradient in the direction perpendicular to both, the applied magnetic field and the thermal flux, was observed. This observation suggests existence of magneto-acoustic phenomena in TGG. One of them, the acoustic Farady effect, has been observed and is reported here (see also Ref. [\onlinecite{conf}]).

\begin{figure}
	\includegraphics[width=\linewidth, keepaspectratio]{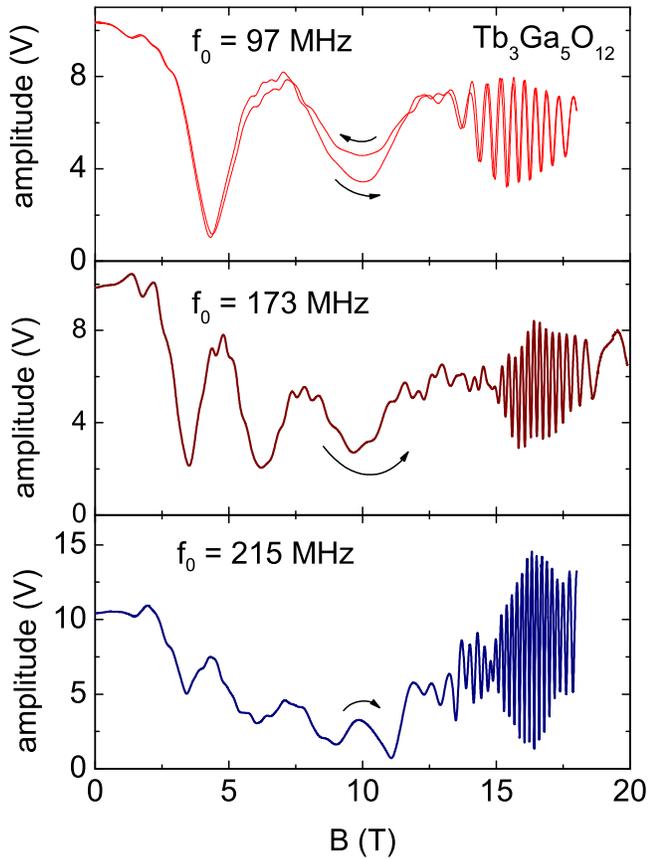} 
  \caption{(Color online) Amplitude oscillations of the acoustic $c_{44}$ mode versus static magnetic field in Faraday geometry  $B \parallel k \parallel [100]$, $u \parallel [010]$ measured at frequency of 97, 173 and 215 MHz and constant temperature of \unit{1.4}{\kelvin}. The arrows indicate the field sweep direction.}
	\label{ampl_stat}
\end{figure}

The intriguing magneto-elastic properties of TGG can be described using the magneto-elastic interaction of the form
\begin{equation}
    H = \sum_{\Gamma} g_{\Gamma} \epsilon_{\Gamma} O_{\Gamma},
\label{me}    
\end{equation}
which represents the basic interaction mechanism of the phonon modes with the orbital moments of the Tb$^{3+}$ ions, where $\Gamma$ denotes the symmetry label, $\epsilon_{\Gamma}$ - the strain components, $O_{\Gamma}$ - the quadrupolar operators of the Tb ion, and $g_{\Gamma}$ - the magneto-elastic coupling constant. 

The elastic constants $c_{\rm 11}$, $(c_{\rm 11}-c_{\rm 12})/2$, and $c_{\rm 44}$ in TGG exhibit strong anomalies below \unit{100}{\kelvin}. \cite{araki} These anomalies are clear evidence for the CEF magneto-elastic interaction described by Eq. (1). The large coupling constant for the $c_{\rm 44}$ mode $g(\Gamma_5)$ points to the possibility for investigating special magneto-acoustic effects, among others, the acoustic Faraday effect. This effect was also observed previously in ferrimagnetic Y$_3$Fe$_5$O$_{12}$ (YIG), in antiferromagnets, and paramagnets. For a review see Ref. [\onlinecite{luthi}]. 

In analogy to the optical Faraday effect, for the magneto-acoustic Faraday effect a transverse acoustic wave propagating along a fourfold cubic axis has two, left and right circularly polarized, components. In an applied magnetic field these components have different velocities. This leads to a rotation of the linear polarization vector as a function of magnetic field, seen as oscillations in amplitude of the acoustic signal. For the optical Faraday effect, the rotation angle per unit length is given by $\Phi/L = VB$, with $V$ being the Verdet constant. We will show that for the magneto-acoustic Faraday effect, $\Phi/L$ can be more complicated.

\section{Experimental results}

The experiments have been performed on a TGG single crystal oriented for propagating the sound wave with wave vector {\bf k} along the [100]-direction. The sample length along the direction of the  sound-wave propagation was \unit{4.005}{mm}. The ultrasonic attenuation and velocity have been measured with a set-up as described at great length in Ref. [\onlinecite{luthi}, \onlinecite{physB}]. LiNbO$_3$ transducers have been used with polarization {\bf u} along the [010] direction. Fields up to \unit{20}{\tesla} have been provided by a commercial superconducting magnet system. Experiments beyond that field have been performed in pulsed magnets. Both systems have been equipped with $^4$He-flow cryostats.

\subsection*{Static fields}

In static magnetic fields, we have performed measurements for frequencies 97, 173, 177, 215, and \unit{337}{\mega \hertz} at constant temperature of \unit{1.4}{\kelvin}. Figure \ref{ampl_stat} shows typical patterns of the magneto-acoustic Faraday oscillations observed for TGG at different frequencies. The amplitude is plotted versus field. For the fields above about \unit{13}{T} the transducer detects subsequent maxima for  polarization angles $\Phi = n\pi$ and subsequent minima for $\Phi = (2n+1)\pi/2$ in intensity or amplitude. Near \unit{20}{T} the oscillations cease. In pulsed-field experiments we have observed that the oscillations reappear above \unit{20}{\tesla} (see next subsection).

\begin{figure}
	\includegraphics[width=\linewidth, keepaspectratio]{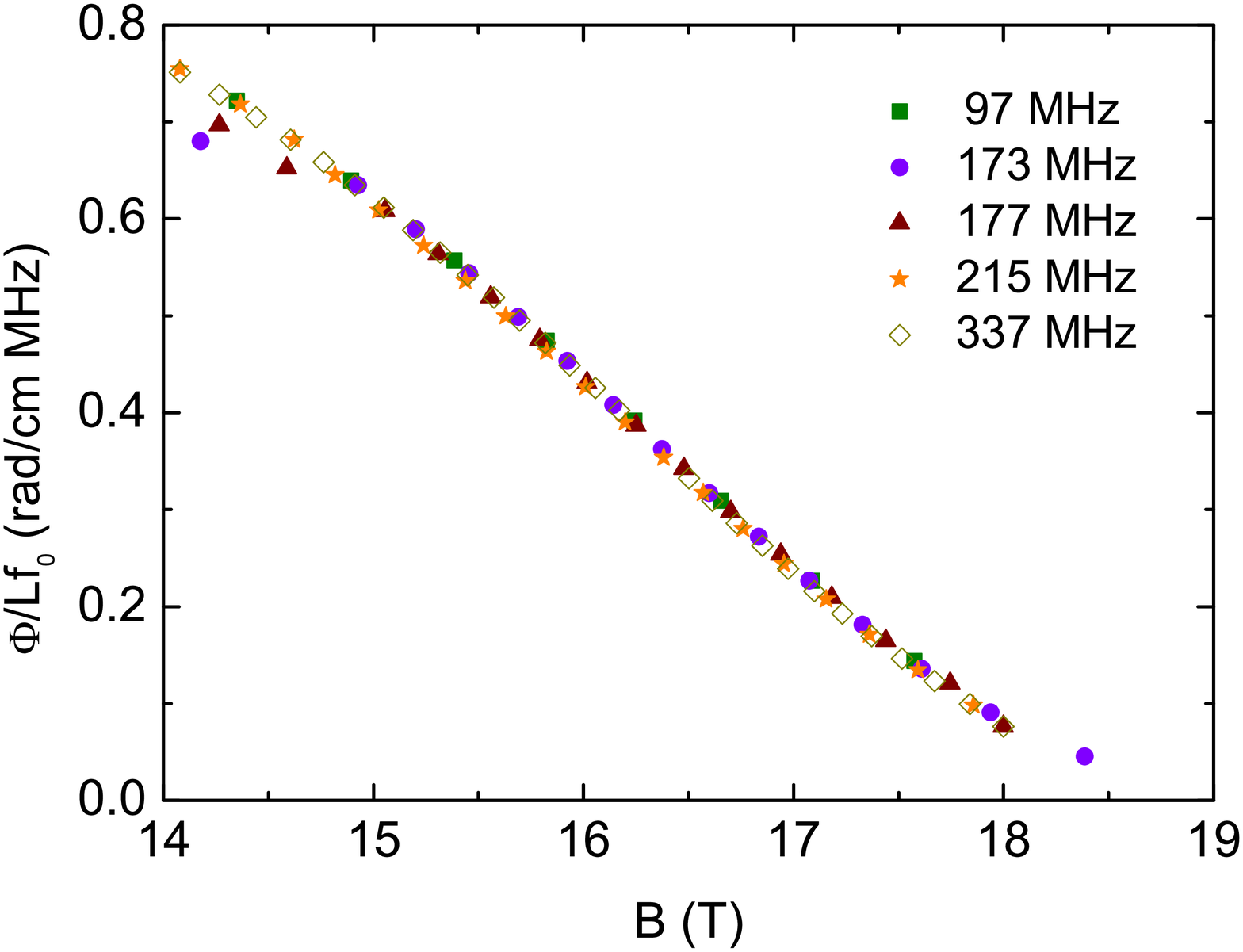}
 	\caption{(Color online) Rotation angle per unit length and frequency $\Phi/L f_0$ as a function of magnetic field for frequencies 97, 173, 177, 215 and 337 MHz. $\Phi$ was taken from the amplitude maxima from Fig. \ref{ampl_stat}.} 
 	\label{rotanglestat}

\vspace{.5 cm}

  \includegraphics[width=\linewidth, keepaspectratio]{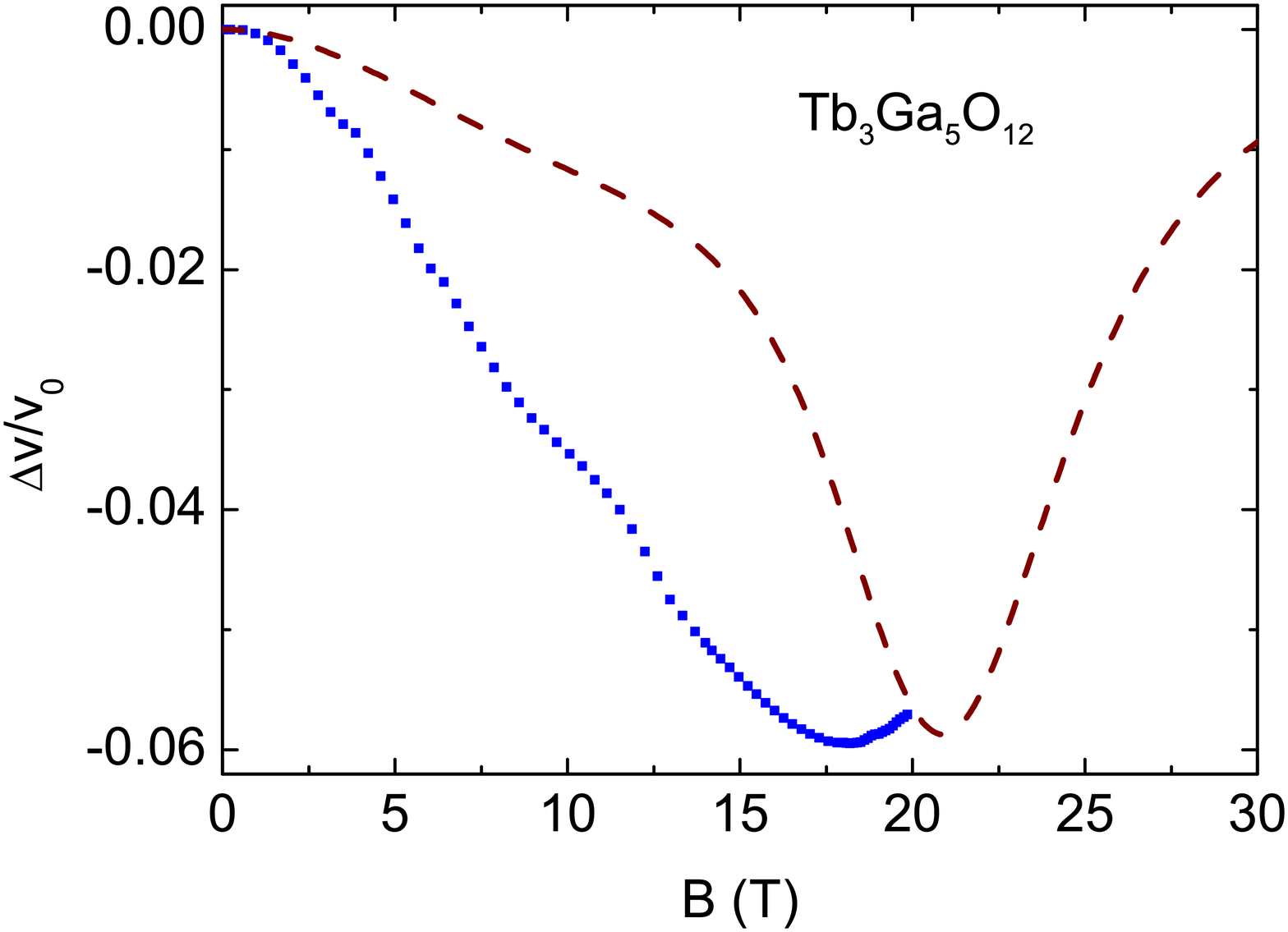}
	\caption{(Color online) The relative velocity change for the acoustic $c_{44}$ mode as a function of magnetic field for $T = 1.4$ K, ${\bf B} \parallel {\bf k} \parallel [100]$, polarization vector ${\bf u} \parallel [010]$. Symbols - experimental data, dashed line - theoretical results.}
	\label{statvelocity}
\end{figure}

In Fig. \ref{rotanglestat}, we plot the rotation angle per unit length and frequency $\Phi/L f_0$, where $L$ is the acoustic path length for a particular echo (sample length in our case) and $f_0$ is the  measurements frequency. Presuming a linear frequency dependence, the $\Phi/L$ data for different frequencies have been divided by the measurements frequency $f_0$. 
It is seen that indeed the linear frequency dependence is fulfilled to a high degree of accuracy. Such a linear dependence was observed before in the paramagnetic state of CeAl$_2$. \cite{ceal} This result is very astonishing because earlier theoretical treatments of the magneto-acoustic Faraday effect, for which only the magneto-elastic coupling of Tb 4$f$ electrons with acoustic phonons has been taken into account, show always a quadratic frequency dependence for $B \gg 2 \pi f_0/\gamma$ with $\gamma$ being the gyromagnetic ratio.\cite{tuck, thalm} This condition is easily obeyed for $B > \unit{10}{\tesla}$ and frequencies in the \unit{100}{\mega\hertz} range. For \unit{100}{\mega\hertz}, the cyclic frequency $\omega = 2 \pi f_0$, and the ratio $\omega/\gamma = \unit{3.7}{\milli\tesla}$. This discrepancy was resolved recently \cite{thalm_new} and is discussed later in the theory section.

In addition, we measured the field dependence of the velocity of the $c_{\rm 44}$ mode at fixed temperature of 1.4 K (Fig. \ref{statvelocity}). This mode exhibits a strong softening with a pronounced minimum at about \unit{18}{\tesla}, i.e., at the field where the lowest level of the quasi-triplet mode crosses the upper doublet mode (see Section 3). Here the maximum softening of the sound velocity amounts to $\Delta v/v_0 = 6 \%$. 

We describe $\Delta v/v_0$ by calculating the strain susceptibility in the magnetic field  using the same quadrupolar operator $O_{\Gamma} = O_{xy} + 0.6 O_2^0$ as for the temperature dependence of $c_{\rm 44}$.\cite{araki, low} We use the formula 

\begin{eqnarray*}
c_{\Gamma} = c_{\Gamma}^{0}(B=0) -\frac{2}{3} g_{\Gamma}^2 N\chi_{\Gamma}
\end{eqnarray*}
with measured elastic constant $c_{\Gamma}^0 = \rho v_0^2 = 9.67 \times \unit{10^{10}}{\joule/\meter^3}$. Here $\rho$ is the mass density. Further, $N=1.28 \times \unit{10^{22}}{\centi \meter^{-3}}$ is the number of Tb-ions per cm$^3$, and $\chi_{\Gamma}$ is the strain susceptibility, which is defined in analogy to the magnetic susceptibility as $\chi_\Gamma = d\langle O_\Gamma \rangle /d \epsilon_\Gamma$. 

There is a rough agreement of the overall shape of the calculated $\Delta v/v_0 $ (dashed line)
with the experimental curve (solid line), though the calculated minimum at \unit{21}{\tesla} is at slightly higher field than the experimentally observed minimum in static fields. 
As for the coupling constant we surprisingly  find  a value of $g_{\Gamma} \approx \unit{21}{\kelvin}$, which is roughly a factor of two smaller than the coupling  $g_{\Gamma}=\unit{45.5}{\kelvin}$ determined  from a fit of $c_{\rm 44}(T)$.\cite{low}

We tentatively attribute this reduction to the fact, that the CEF levels
display a nonzero (possibly field-dependent) linewidth, which is not
included in our calculation. Also we expect that a careful averaging over all inequivalent ions, which will force us to incorporate new strain coupling constants, will  modify the picture to a certain degree. In the above calculation we took this averaging into account by an overall factor $\frac{2}{3}$, because we found $\chi_{\Gamma}(T)$ to be strongly suppressed for one third of the ions. Details of the calculation and possible effects of the averaging procedure will be presented in a forthcoming publication. \cite{low}

\subsection*{Pulsed fields}

\begin{figure}
	\includegraphics[width=\linewidth, keepaspectratio]{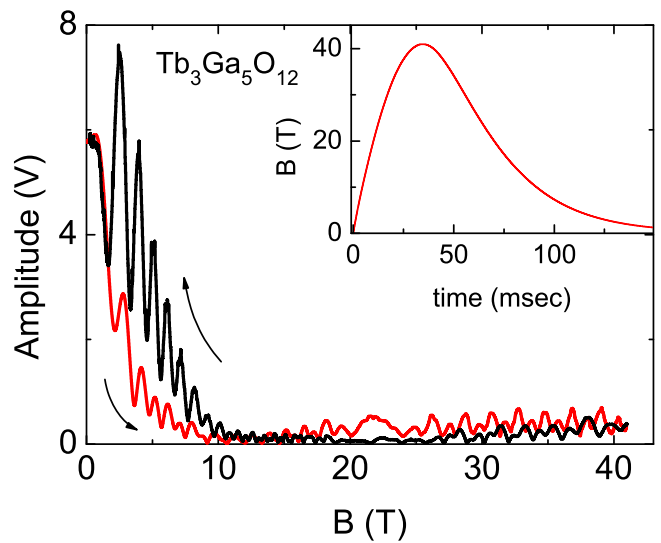} 
  \caption{(Color online) Amplitude oscillations of the acoustic $c_{\rm 44}$ mode versus pulsed magnetic field in Faraday geometry  ${\bf B} \parallel {\bf k} \parallel [100]$, ${\bf u} \parallel [010]$ measured at frequency of 93 MHz and temperature of 1.4 K. The arrows indicate the field sweep direction. Inset: time profile of pulsed magnetic field.}
	\label{ampl_pulse}
\end{figure}

\begin{figure}[h]
  \includegraphics[width=\linewidth, keepaspectratio]{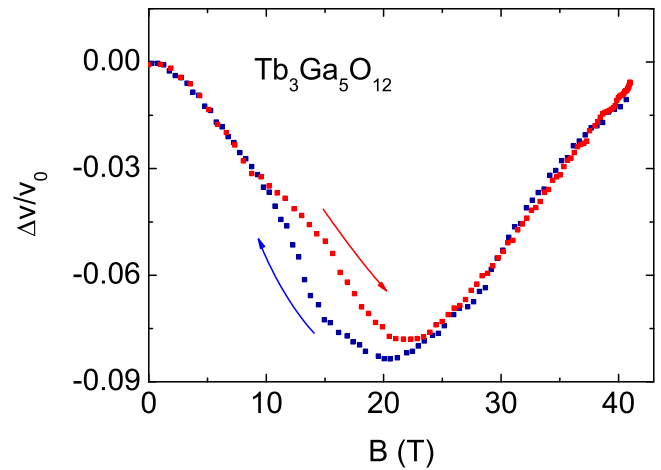}
	\caption{(Color online) The relative velocity change for the acoustic $c_{\rm 44}$ mode as a function of magnetic field for $T = 1.4$ K, ${\bf B} \parallel {\bf k} \parallel [100]$, polarization vector ${\bf u} \parallel [010]$. Arrows indicate direction of the field sweep.}
	\label{velocity}
\end{figure}

To confirm the reappearance of the acoustic Faraday oscillations above \unit{20}{\tesla} we further performed pulsed-field experiments.
In Figs. \ref{ampl_pulse} and \ref{velocity}, we show the field dependence of the sound amplitude oscillations and sound velocity taken at \unit{93}{\mega\hertz}. The data demonstrate some hysteresis.
The acoustic Faraday oscillations do occur both below and above \unit{20}{\tesla}.

\begin{figure}
	\includegraphics[width=\linewidth, keepaspectratio]{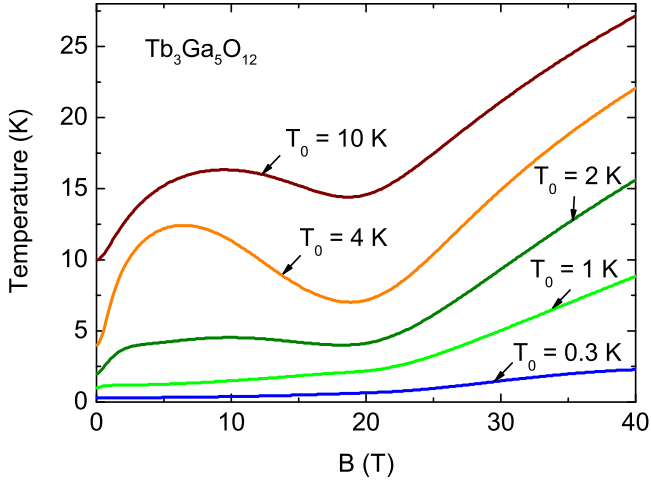}
  \caption{(Color online) Temperature change due to the MCE for ${\bf B} \parallel [100]$ calculated for  different initial temperatures of $T_0 = 0.3$, $1$, $2$, $4$ and $10$ K.}
	\label{mce_temp}
\end{figure}

\begin{figure}
	\includegraphics[width=\linewidth, keepaspectratio]{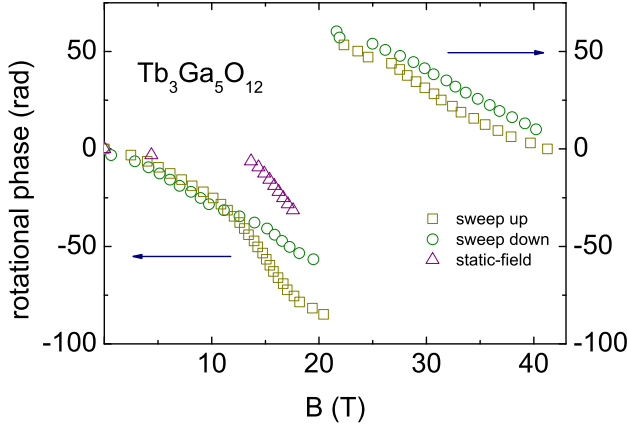}
	\caption{(Color online) Rotation angle $\Phi$, extracted from data in Figs. \ref{ampl_pulse} and \ref{ampl_stat} for pulsed and static fields, respectively. Ultrasound frequency is 93 MHz in pulsed-field measurements; 95 MHz - in static-field measurements.}
	\label{rotanglepulse}
\end{figure}

However, some puzzling phenomena appear in pulsed fields. The oscillations in Fig. \ref{ampl_pulse} can be observed in the complete field range with the exception of a damped region around \unit{20}{\tesla}. This is in sharp  contrast to the static-field measurements shown in Fig. \ref{ampl_stat}, where the oscillation period is very large for  $B <$ \unit{13}{\tesla}. In addition, the decrease of the corresponding sound velocity is \unit{30}{\%} stronger in pulsed fields (Fig. \ref{velocity}) than in static fields (Fig. \ref{statvelocity}). These differences between static- and pulsed-field data possibly are due to a nonequilibrium situation during the pulsed-field experiment. The time dependence of $B$ is given in the inset of Fig. \ref{ampl_pulse}. For the static field measurements we swept the field at a rate of \unit{75}{\milli\tesla/\minute}.

Under quasi-adiabatic conditions, a huge magnetocaloric effect (MCE) can occur in pulsed fields. In TGG, the MCE effect was investigated theoretically for fields applied in different crystallographic directions and experimentally for  $B \parallel [110]$. \cite{mce, levitin} We performed calculations of the MCE as in Ref. [\onlinecite{mce}], averaging over the CEF levels of the three Tb${^3+}$ ions in inequivalent positions. In Fig. \ref{mce_temp} we present our results for different starting temperatures and for magnetic field applied along the [100] direction. From our calculations we expect nonmonotonic temperature changes with magnetic field, as seen in Fig. \ref{mce_temp}. The MCE, together with thermal relaxation to the bath may lead to complicated hysteretical heating and subsequent cooling effects. This could at least partially explain the increased amplitudes of Faraday oscillations in the descending field. A quantitative analysis of these effects is challenging and stays outside the scope of this work.  

\begin{figure}
\includegraphics[width=\linewidth, keepaspectratio]{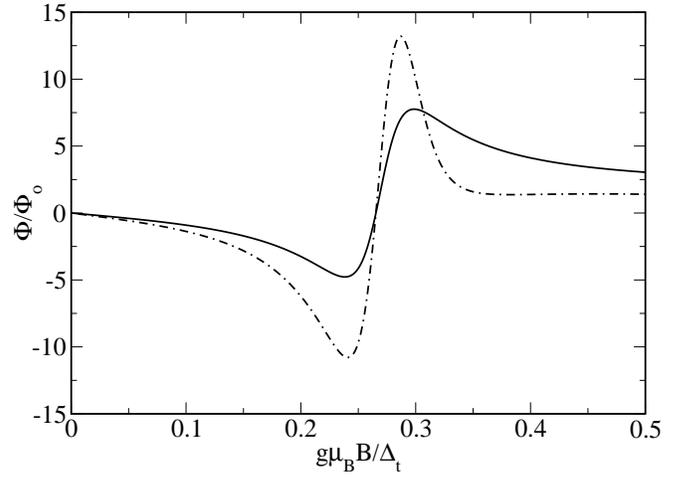}
\caption{The calculated normalised Faraday rotation angle 
$\Phi/\Phi_0 (\Phi_0=\frac{1}{2}k(m_Q\tilde{E}\tilde{g})^2\omega_o; k=\omega/v)$
as  function of magnetic field for $T\ll\Delta_t$, ${{\bf B}  \parallel {\bf k} \parallel [001]\perp {\bf u}}$.
Full line: using bare a-o coupling $\tilde{E}$; broken line: using renormalized $\tilde{E}_t$ with 
$2m_Q^2\tilde{g}^2/\tilde{E} = 0.05$. $m_Q$ is the quadrupolar ($O_{xz},O_{yz}$) singlet- triplet
matrix element for $\Delta_t^\pm$ excitations. Finite linewidths for $\omega_o$ and $\Delta_t^\pm$ have
been used.}
\label{farcubic}
\end{figure}

From the acoustic amplitude oscillations of the pulsed field data the rotational phase versus magnetic field can be extracted. This has been done in Fig. \ref{rotanglepulse} for oscillations in the field range B $<$ \unit{20}{\tesla} and B $>$ \unit{20}{\tesla}. This figure qualitatively agrees with the theoretical predictions, shown in Fig. \ref{farcubic}.

The obtained results for the magneto-acoustic Faraday oscillations and the sound velocity-change are  a good illustration that various side effects have to be taken into account in pulsed-field experiments.

\section{Theory}

In TGG, the cubic crystal leads to fourfold symmetry axes [001] {\it etc}. Therefore, doubly degenerate transverse acoustic (a) and optical (o) lattice vibration (E) modes exist for a wave vector {\bf k} parallel to these axes. Equivalently, they may be presented by left (L) and right (R) circularly polarized modes, which are complex conjugates and degenerate due to time-reversal symmetry. Application of a magnetic field breaks this symmetry and leads to different wave numbers for L and R modes of a given frequency $\omega$, provided that there are magnetic degrees of freedom, which couple sufficiently strong to the lattice vibrations. The former may be collective spin waves as well as localized CEF excitations. In the case of spin waves in a ferro (ferri-) magnet the simplest theoretical treatment of the magneto-acoustic Faraday effect takes into account the coupled equation of motion for the classical magnetic moments {\bf M} and the acoustic displacements {\bf u}. Then  one obtains a Faraday-rotation angle per unit length given by the difference of the L and R wave numbers as \cite{wang,luthi}
\begin{figure}
	\includegraphics[width=\linewidth, keepaspectratio]{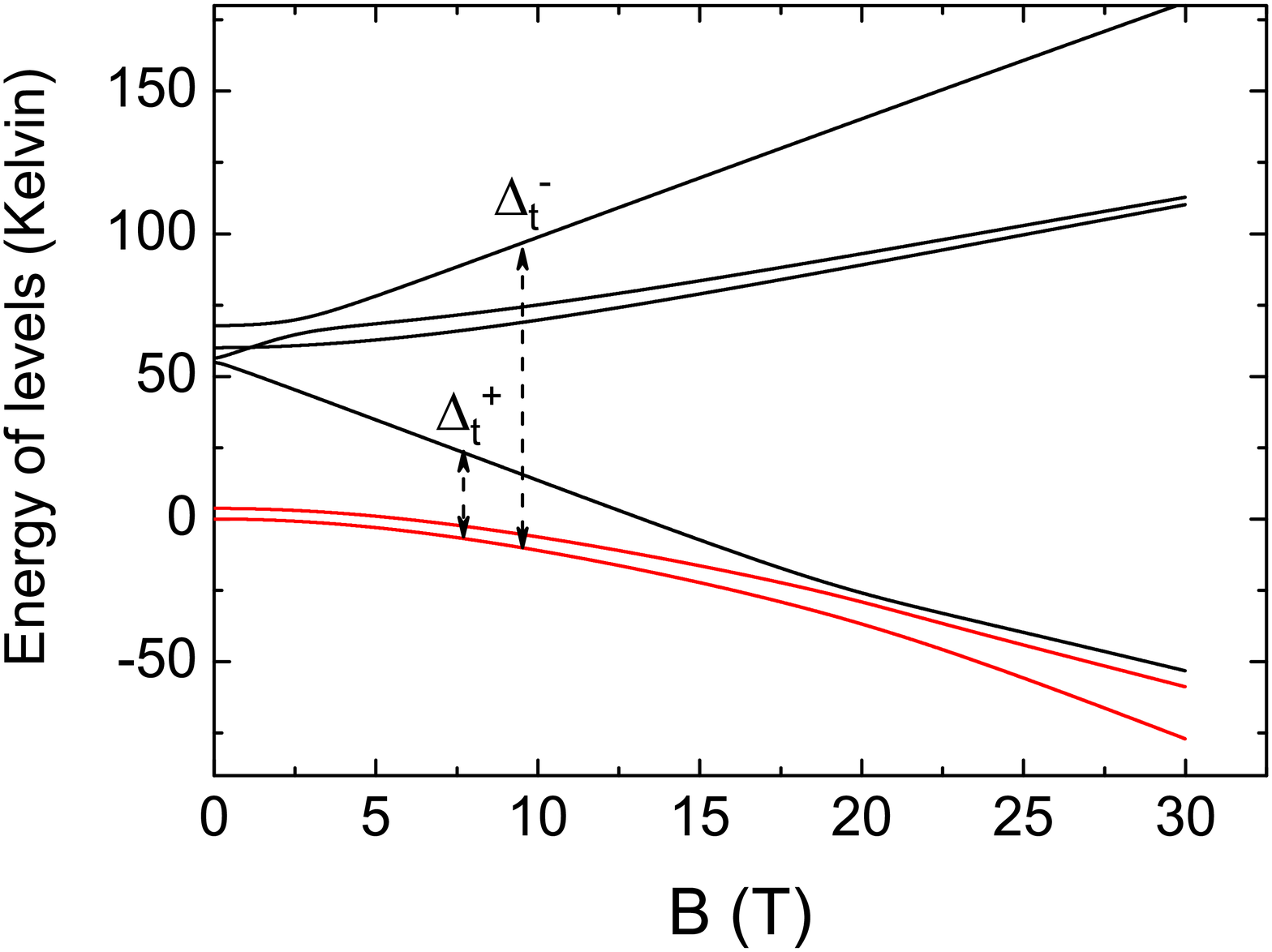} 
  \caption{(Color online) Lowest CEF energy levels for field applied in [100] direction. Notice the crossing of the lowest triplet level with the upper doublet level.}
  \label{CEFlevels}
\end{figure}
\begin{equation}
\frac{\Phi}{L} = \frac{1}{2}(k_L - k_R) = \frac{1}{2}sk \frac{\omega}{[\omega^2 - (\omega_m - s)^2]}.
\label{FRferro}
\end{equation}
Here, $\omega_m = \gamma B_{\rm eff}$ and $s = \gamma b^2 / (v^2M_0)$, where $B_{\rm eff}$ is the effective molecular field, $M_0$ stands for the magnetic moment in the propagation direction and $\gamma, b$ and $v$ are the gyromagnetic ratio, the magnetoelastic coupling constant and the sound velocity, respectively. For large enough fields $\omega_m\gg \omega$ and then $\Phi/L \propto \omega^2$, since $k=\frac{\omega}{v}$. Calculations for $S = 1/2$ paramagnets \cite{tuck}  and multilevel paramagnetic CEF systems \cite{thalm} also give a dependence  $\Phi/L \propto \omega^2$. Thus, existing theories so far consistently predicted a different frequency dependence than the linear frequency dependence of $\Phi/L$ observed either in (paramagnetic) CeAl$_2$ \cite{ceal} or now in TGG. The Faraday rotation in magnetically ordered compounds such as YIG \cite{Matthews62,Guermeur67} and paramagnets with magnetic impurities \cite{Guermeur68} was only measured for a single frequency. We suspect that also in these cases the discrepancy to the quadratic frequency dependence of Eq.~(\ref{FRferro}) exists. In this equation, the prefactor $k =\frac{2\pi}{\lambda}$ is unavoidable because it counts the number of sound-wave periods per unit length. The problem lies with the factor $\omega$ in the numerator of Eq.~(\ref{FRferro}) which stems from the 
nondiagonal $(xy)$ and purely imaginary component  of the dynamic magnetic susceptibility that has to vanish for $\omega\rightarrow 0$, i.e., the splitting of the acoustic L,R modes has to 	approach zero for $k\rightarrow 0$. Thus, it seems that any mechanism involving only acoustic phonons and magnetic excitations will lead to the  $\Phi/L\propto \omega^2$ frequency dependence in contrast to experiment. 
A solution for this discrepancy was suggested recently. \cite{thalm_new} In non-Bravais lattices such as TGG (and also CeAl$_2$) ${\bf k}\rightarrow 0$ optical phonons may have a direct influence on sound waves due to a coupling of elastic long-wave length strains and internal (optical) displacements, provided they have the same symmetry. For TGG, there are many optical phonons of the right (doubly degenerate) E symmetry \cite{Papagelis02} that can couple to the E-type transverse acoustic strain components. \cite{thalm_new} The acoustic phonons have a direct magnetoelastic coupling to CEF transitions (Eq.~(\ref{me})), which again leads to a Faraday rotation with quadratic frequency dependence of the generalized form of Eq.~(\ref{FRferro}).\cite{thalm} Furthermore, the optical phonons also couple
to the CEF excitations. This leads to a splitting of optical E-type (transverse) phonons with bare frequency $\omega_o$ into circular polarized modes with split frequencies $\tilde{\omega}^{\pm 2}_o$. Via the effective coupling of acoustic and optical modes this creates an indirect contribution to the Faraday rotation of the form
\begin{eqnarray}
\frac{\Phi}{L}=\frac{1}{2}k\tilde{E}_t^2\frac{\omega_o^2}{2}
\Bigl(\frac{1}{\tilde{\omega}_o^{-2}}-\frac{1}{\tilde{\omega}_o^{+2}}\Bigr),
\label{Phiao}
\end{eqnarray}
with circular optical phonon frequencies and  effective a-o coupling $\tilde{E}_t$ given by
\begin{eqnarray}
\tilde{\omega}_o^{\pm 2}&=&\omega_o^2(1+\tilde{g}_o^2\la\la\hat{O}_{xz}\hat{O}_{xz}\ra\ra'_{\omega_o}
\pm\tilde{g}_o^2\la\la\hat{O}_{xz}\hat{O}_{yz}\ra\ra''_{\omega_o})\nonumber\\
\tilde{E}_t&=&\tilde{E}+\tilde{g}\la\la\hat{O}_{xz}\hat{O}_{xz}\ra\ra'_0
\label{quadsus}
\end{eqnarray}
Here, $\tilde{g}_a, \tilde{g}_o, \tilde{g}=(\tilde{g}_a\tilde{g}_o)^\frac{1}{2}$ are acoustic, optical, and mixed magnetoelastic coupling constants and $\tilde{E}$ is the bare a-o coupling of phonons. Furthermore, the double brackets denote the dynamical local susceptibilities (prime: real part, double prime: imaginary part, subscript: frequency) of 4f quadrupolar-moment operators $O_{xz}=(J_xJ_z+J_zJ_x)$ and $O_{yz}=(J_yJ_z+J_zJ_y)$ which couple to x, y polarized acoustic and optical phonons.

For ${\bf k} \rightarrow 0$ in Eq.~(\ref{Phiao}), the splitting of optical phonon modes $\tilde{\omega}_o^\pm$  stays finite and, therefore, ($k=\omega/v$) we have a linear frequency dependence $\Phi/L\sim \omega$ for the indirect Faraday-rotation angle induced via a-o phonon coupling. It has been argued \cite{thalm_new} that for reasonable parameters this term dominates the quadratic term in Eq.~(\ref{FRferro}). 

For the explicit calculation of the contribution linear in $\omega$ we need the above quadrupolar susceptibilities for the CEF level scheme of TGG in a magnetic field. Due to the low orthorhombic symmetry the $J=6$ manifold splits into 13 singlets \cite{guillot} at three inequivalent Tb$^{3+}$ sites. The Zeeman splitting of the lowest levels of the most relevant inequivalent Tb$^{3+}$ ion for ${\bf B} \parallel [100]$  is shown in Fig.~\ref{CEFlevels}. Apparently, some quasi-degeneracies remain for zero field, e.g. an excited magnetic quasi-triplet at $\Delta_t\simeq 57$ K. Therefore, the low-lying levels may be approximately described by a simplified CEF model with cubic symmetry as proposed in Ref.~\onlinecite{araki}. It is  used here for the calculation of the quadrupolar susceptibilities in  Eq.~(\ref{quadsus}). Their explicit expressions for the cubic doublet (ground state) - triplet (excited state) model are given in Ref.~[\onlinecite{thalm_new}]. The result of a model calculation for the Faraday rotation using  Eq.~(\ref{quadsus}) and Eq. (\ref{Phiao}) is shown in Fig.~\ref{farcubic}.

It is qualitatively very similar to the result of the pulsed-field experiment in Fig.~\ref{rotanglepulse}. The main CEF transitions contributing to the Faraday rotation in the cubic model are those between the ground state and the two linearly split (quasi-triplet) excitations with energies $\Delta_t^\pm$  indicated in Fig.~\ref{CEFlevels}. While $\Delta_t^+$ contributes to the effective a-o interaction $\tilde{E}_t$, $\Delta_t^-$ leads mainly to the splitting of optical $\tilde{\omega}_o^\pm$ modes. The Faraday rotation develops a resonance-like behavior due to two effects: The ground state - triplet level crossing ($\Delta_t^+(B)=0$) and the resonance with the optical phonon frequency  ($\Delta_t^-(B)=\omega_o$) which happen at approximately the same field of $\unit{17-20}{\tesla}$.

\section{Summary}

In summary, we have observed and analyzed the magneto-acoustic Faraday effect in TGG in magnetic fields up to \unit{42}{\tesla}. A strong damping of the Faraday oscillation amplitude has been found in the region of the CEF level crossing at about \unit{20}{\tesla}. We have also shown that the rotation angle of the shear-wave polarization vector is strictly linear in the sound-wave frequency. The strong softening of the acoustic $c_{{\rm 44}}$ mode as a function of magnetic field has been detected and reproduced quantitatively in a calculation taking into account magneto-elastic interaction. For the acoustic Faraday rotation our theoretical model is based on magnetoelastic coupling of 4$f$ electrons to both, acoustic and optical phonons and an effective interaction between them. The finite splitting of optical phonons for long wavelengths and the acoustic-to-optical phonon coupling results in a contribution to the Faraday-rotation angle which is linear in the sound-wave frequency \cite{thalm_new} as it was observed in the experiments.

\begin{acknowledgments}
This work has been partially supported by the Euromagnet II research program under EU contract 228043. We also cordially acknowledge the useful discussions with Prof. Dr.~W.~Weber.

\end{acknowledgments}

\end{document}